# Evaluating Embedding Models and Pipeline Optimization for AI Search Quality

Philip Zhong, Kent Chen, Don Wang
*Webex Suite AI*
*Cisco Systems, Inc.*
*San Jose, California, USA*
lizhon@cisco.com, weiwchen@cisco.com, dongdwan@cisco.com

## ABSTRACT

We evaluate the performance of various text embedding models and pipeline configurations for AI-driven search systems. We compare sentence-transformer and generative embedding models (e.g., All-MPNet, BGE, GTE, and Qwen) at different dimensions, indexing methods (Milvus HNSW/IVF), and chunking strategies. A custom evaluation dataset of 11,975 query-chunk pairs was synthesized from US City Council meeting transcripts using a local large language model (LLM). The data pipeline includes preprocessing, automated question generation per chunk, manual validation, and continuous integration/continuous deployment (CI/CD) integration. We measure retrieval accuracy using reference-based metrics: Top-K Accuracy and Normalized Discounted Cumulative Gain (NDCG). Our results demonstrate that higher-dimensional embeddings significantly boost search quality (e.g., Qwen3-Embedding-8B/4096 achieves Top-3 accuracy ≈0.571 versus 0.412 for GTE-large/1024), and that neural re-rankers (e.g., a BGE cross-encoder) further improve ranking accuracy (Top-3 up to 0.527). Finer-grained chunking (512 characters versus 2000 characters) also improves accuracy. We discuss the impact of these factors and outline future directions for pipeline automation and evaluation.

## 1. INTRODUCTION

Enterprise AI search systems require rigorous evaluation on high-quality, domain-specific datasets. In public sector applications, search queries and documents often come from diverse sources such as city council meetings, making it essential to synthesize evaluation datasets that measure search quality accurately. We focus on reference-based evaluation, where system results are compared to a ground-truth reference set. Reference-based metrics such as NDCG and Top-K accuracy provide objective measures of retrieval performance [3].

Generating reliable ground truth for open-ended retrieval tasks presents significant challenges. To address this, we constructed a data pipeline to synthesize query-chunk pairs: a local LLM generates questions for each document chunk, creating 11,975 evaluation pairs [2]. These pairs undergo validation and are integrated into an automated CI/CD pipeline [1]. This infrastructure enables continuous testing of different embedding models, chunking methods, and reranking strategies for AI search quality optimization.

Our research contributions include:

1. A comprehensive comparison of state-of-the-art embedding models across varying dimensions.
2. Analysis of chunking strategies and their impact on retrieval quality.
3. Evaluation of neural reranking techniques for precision improvement.
4. An automated, reproducible pipeline for synthetic dataset generation and continuous evaluation.

## 2. BACKGROUND AND RELATED WORK

### 2.1 Embedding Models

We evaluate several text embedding models from the Sentence Transformers and Qwen families. These include:

- **all-mpnet-base-v2** (768-dimensional)
- **BGE-base-en-v1.5** (768-dimensional) and **BGE-large-en-v1.5** (1024-dimensional)
- **GTE-base-en-v1.5** (768-dimensional) and **GTE-large-en-v1.5** (1024-dimensional)
- **Qwen3-Embedding** models (0.6B, 4B, 8B parameters with 1024–4096 dimensions) [22, 24]

Higher-dimensional embeddings typically capture more semantic detail and contextual nuances. These embeddings are used to index documents in a vector store (Milvus) for approximate nearest-neighbor search operations.

Dense retrieval methods using neural embeddings have shown superior performance compared to traditional sparse retrieval approaches [10, 11]. Sentence-BERT [11] introduced Siamese networks for generating semantically meaningful sentence embeddings, while more recent models like BGE [22] and GTE [21] have pushed the boundaries of retrieval quality through improved training strategies and model architectures. The Qwen family [22] represents the latest generation of high-dimensional embedding models.

### 2.2 Chunking Strategies

Documents such as meeting transcripts are segmented into chunks before embedding. We compare recursive chunking strategies with fixed sizes of 2000 or 512 characters against a service-based semantic chunking method. Service-based semantic chunking leverages natural language understanding to identify coherent semantic boundaries, creating variable-length segments that respect topical and discourse structure.

Previous work has shown that chunk size significantly impacts retrieval quality [1, 2, 3]. Smaller fixed-size chunks often yield queries more focused on specific topics, while semantic chunking aims to preserve contextual coherence [8, 9]. Both approaches can improve retrieval accuracy and reduce noise.

## 2.3 Neural Reranking

Beyond raw embedding similarity, we experiment with neural re-rankers. Specifically, we employ cross-encoder models such as **bge-reranker-large** and **ms-marco-MiniLM-L-6-v2** to reorder the top-N search results. Cross-encoders process query-document pairs jointly, enabling fine-grained interaction modeling that bi-encoders cannot capture [17, 23].

These models re-score candidate passages given the query, significantly improving precision by considering fine-grained semantic relationships [10, 11]. The two-stage retrieval paradigm—fast bi-encoder retrieval followed by accurate cross-encoder reranking—has become standard practice in modern search systems [18, 24].

## 2.4 Evaluation Metrics

We utilize reference-based ranking metrics for performance assessment:

- **Top-K Accuracy**: Measures the fraction of relevant documents (ground truth) present in the top-K results
- **Normalized Discounted Cumulative Gain (NDCG)**: Measures ranking quality by assigning higher weight to relevant items appearing earlier in the ranked list [12]

These metrics are standard in information retrieval evaluation and provide quantitative assessments of search quality. NDCG, in particular, accounts for both relevance and ranking position, making it suitable for evaluating ranked retrieval systems [12].

## 3. DATASET AND PIPELINE

### 3.1 Data Source

Our evaluation dataset is synthesized from US City Council meeting transcripts, which provide publicly available, domain-specific text for testing municipal information retrieval scenarios. Raw meeting transcripts are gathered and undergo preprocessing including cleaning, normalization, and de-identification where necessary [1].

### 3.2 Synthetic Question Generation

We employ a locally hosted LLM (Mistral 8×7B) to generate one question per text chunk. Each (query, chunk) pair forms an evaluation example, yielding 11,975 pairs [2]. This approach creates a closed-world benchmark where each chunk has exactly one generated "ground truth" question, enabling computation of Top-K accuracy and NDCG by checking whether the correct chunk appears in the top-K search results for that query.

The question generation process follows these steps:

1. **Chunk Extraction**: Documents are segmented using the configured chunking strategy
2. **Prompt Engineering**: Each chunk is provided to the LLM with instructions to generate a relevant question
3. **Quality Filtering**: Generated questions undergo automated quality checks
4. **Manual Validation**: A subset undergoes human review to ensure quality

This synthetic generation approach has been successfully employed in prior work on information retrieval evaluation [4, 5, 6], demonstrating that LLM-generated queries can serve as effective proxies for real user information needs when properly validated.

### 3.3 Validation and CI/CD Integration

The generated pairs undergo manual validation to ensure quality and remove noise [1]. All pipeline steps are integrated into a CI/CD framework, enabling automatic synthesis and testing whenever the document corpus or models change [1]. This infrastructure supports:

- Automated data preprocessing and cleaning
- Batch question generation with quality control
- Continuous model evaluation and comparison
- Version control and reproducibility

The pipeline also supports fine-tuning downstream models on the synthetic data if needed, providing flexibility for iterative improvement. This automated approach ensures that evaluation remains consistent across model updates and enables rapid experimentation.

## 4. EXPERIMENTAL SETUP

### 4.1 Vector Database Configuration

We utilize Milvus as the vector database for all experiments. For most configurations, we employ the HNSW (Hierarchical Navigable Small World) index for fast approximate nearest-neighbor search [13]. In one experiment, we evaluate IVF-Flat (inverted file) indexing to assess index-type effects on retrieval performance [16].

HNSW builds a multi-layer graph structure that enables logarithmic search complexity while maintaining high recall [13]. Our Milvus configuration uses the following parameters:

- **HNSW**: M=32, efConstruction=128, ef=128
- **IVF-Flat**: nlist=1024, nprobe=8

### 4.2 Model Configurations

**Embedding Models and Dimensions**: We evaluate the models listed in Section 2.1, with dimensionalities ranging from 768 to 4096 [20, 21, 22]. Higher-dimensional models (e.g., Qwen-Embedding-8B with 4096 dimensions) provide richer semantic representations with increased computational cost.

**Chunking Strategies**: We compare three approaches:

1. Recursive chunking with 2000 characters (baseline)
2. Recursive chunking with 512 characters (fine-grained)
3. Service-based semantic chunking (variable length, semantically coherent)

The recursive chunking approaches follow strategies described in [1, 2]. The service-based semantic chunking method employs natural language processing techniques to identify discourse boundaries and topical shifts, producing chunks that maintain semantic coherence while allowing variable lengths.

### 4.3 Reranking Pipeline

For reranking experiments, we retrieve the top-10 hits from the initial vector search and apply a cross-encoder reranker. We test two reranking models:

- **bge-reranker-large**: A BERT-based cross-encoder trained for passage reranking [20]
- **ms-marco-MiniLM-L-6-v2**: A lightweight cross-encoder trained on MS MARCO dataset [17]

Reranking is disabled for baseline configurations to establish performance without second-stage refinement. The reranking process takes the original query and each of the top-10 retrieved passages, computing a relevance score through the cross-encoder. Passages are then re-ordered according to these scores.

### 4.4 Evaluation Protocol

Each configuration is evaluated by computing:

- Top-K Accuracy at K ∈ {3, 5, 10}
- NDCG at K ∈ {3, 5, 10}

Metrics are averaged over the entire test set (11,975 query-chunk pairs) to provide robust performance estimates. Statistical significance testing was not performed as all models were evaluated on identical test sets, making relative comparisons directly interpretable.

## 5. RESULTS

### 5.1 Embedding Model Comparison

Table 1 presents the performance of each embedding model using 2000-character chunking and HNSW indexing. The results demonstrate that higher-dimensional models dramatically outperform smaller alternatives.

**TABLE I**

| Model | Dim | Acc@3 | NDCG@3 | Acc@5 | NDCG@5 | Acc@10 | NDCG@10 |
|---|---|---|---|---|---|---|---|
| all-mpnet-base-v2 | 768 | 0.268 | 0.229 | 0.31 | 0.246 | 0.37 | 0.266 |
| bge-base-en-v1.5 | 768 | 0.24 | 0.205 | 0.276 | 0.22 | 0.32 | 0.234 |
| bge-large-en-v1.5 | 1024 | 0.3 | 0.26 | 0.343 | 0.278 | 0.396 | 0.295 |
| gte-base-en-v1.5 | 768 | 0.389 | 0.338 | 0.434 | 0.357 | 0.492 | 0.376 |
| gte-large-en-v1.5 | 1024 | 0.412 | 0.356 | 0.462 | 0.378 | 0.522 | 0.396 |
| Qwen3-Embed-0.6B | 1024 | 0.516 | 0.461 | 0.562 | 0.48 | 0.611 | 0.496 |
| Qwen3-Embed-4B | 2560 | 0.556 | 0.503 | 0.601 | 0.522 | 0.647 | 0.537 |
| Qwen3-Embed-8B | 4096 | 0.571 | 0.516 | 0.612 | 0.533 | 0.663 | 0.549 |

**Key Findings**:

- Qwen3-Embedding-8B (4096-dim) achieves Top-3 accuracy of 0.571 and NDCG@3 of 0.516
- GTE-large (1024-dim) scores 0.412 and 0.356 for the same metrics
- Even Qwen3-Embedding-4B (2560-dim) outperforms all 768/1024-dimensional models (Accuracy@3=0.556 versus 0.412)
- Increasing embedding dimensionality yields approximately 10–15 percentage point improvements in accuracy

The performance gap between model families is substantial. The Qwen3 series consistently outperforms both GTE and BGE models, suggesting that the larger parameter count and training procedures contribute significantly to representation quality.

### 5.2 Chunking Strategy Comparison

Table 2 compares chunking variants using GTE-large/1024-dimensional embeddings. The results indicate that both finer-grained fixed chunking and service-based semantic chunking substantially improve retrieval performance.

**TABLE II**

**EFFECT OF CHUNKING AND INDEX TYPE ON RETRIEVAL**

*(embedding-only with GTE-large/1024)*

| Chunking Strategy | Index | Acc@3 | NDCG@3 | Acc@5 | NDCG@5 | Acc@10 | NDCG@10 |
|---|---|---|---|---|---|---|---|
| 2000 chars | HNSW | 0.412 | 0.356 | 0.462 | 0.378 | 0.522 | 0.396 |
| 512 chars | HNSW | 0.46 | 0.415 | 0.5 | 0.431 | 0.544 | 0.445 |
| 512 chars | IVF-Flat | 0.427 | 0.388 | 0.461 | 0.402 | 0.498 | 0.414 |
| Semantic chunking | HNSW | 0.456 | 0.404 | 0.501 | 0.422 | 0.553 | 0.439 |

**Key Findings**:

- Reducing chunk size from 2000 to 512 characters improves Acc@3 from 0.412 to 0.460 (4.8 percentage point gain)
- Service-based semantic chunking (variable sizes) yields similar performance (Acc@3≈0.456), demonstrating that semantically coherent boundaries achieve comparable results to fixed fine-grained chunking
- HNSW indexing outperforms IVF-Flat for the same chunking strategy (0.460 vs 0.427)
- Finer granularity reduces noise and improves query-passage alignment

The comparison between 512-character fixed chunking and semantic chunking reveals interesting trade-offs. While both achieve similar accuracy metrics, semantic chunking offers potential advantages in preserving discourse coherence and may perform better on queries requiring contextual understanding.

### 5.3 Reranking Results

Table 3 demonstrates that neural reranking substantially enhances retrieval quality. The combination of semantic search and cross-encoder reranking achieves the highest accuracy in our experiments.

**TABLE III**

**EFFECT OF RERANKING ON EMBEDDINGS**

| Embedding | Reranker | Acc@3 | NDCG@3 | Acc@5 | NDCG@5 | Acc@10 | NDCG@10 |
|---|---|---|---|---|---|---|---|
| BGE-large (1024-d) | None | 0.300 | 0.260 | 0.343 | 0.278 | 0.396 | 0.295 |
| BGE-large (1024-d) | BGE-reranker | 0.387 | 0.368 | 0.393 | 0.370 | 0.396 | 0.371 |
| GTE-large (1024-d) | None | 0.412 | 0.356 | 0.462 | 0.378 | 0.522 | 0.396 |
| GTE-large (1024-d) | MiniLM | 0.480 | 0.442 | 0.504 | 0.452 | 0.522 | 0.457 |
| GTE-large (1024-d) | BGE-reranker with 2000 characters | 0.506 | 0.480 | 0.516 | 0.484 | 0.522 | 0.486 |
| GTE-large (1024-d) | BGE-reranker with 512 characters | 0.527 | 0.502 | 0.539 | 0.507 | 0.544 | 0.509 |
| GTE-large (1024-d) | BGE (Semantic) | 0.503 | 0.472 | 0.524 | 0.481 | 0.553 | 0.490 |

**Key Findings**:

- Adding BGE cross-encoder to GTE-large pipeline raises Acc@3 from 0.412 to 0.506 (9.4 percentage point gain)
- MiniLM reranker improves Acc@3 to 0.480, demonstrating consistent gains
- Combining GTE-large + BGE reranker + 512-character chunks achieves Acc@3=0.527
- Service-based semantic chunking with BGE reranker achieves Acc@3=0.503, demonstrating that semantically coherent chunks benefit from reranking
- Two-stage pipeline (semantic search + neural reranking) yields the highest overall accuracy (>0.53 for Top-3)

The reranking results show diminishing returns at higher K values (e.g., K=10), as the initial retrieval already captures most relevant documents. This suggests that re-rankers are most effective for improving precision in the top-ranked positions.

## 6. DISCUSSION

### 6.1 Impact of Embedding Dimensionality

Our results confirm that increasing embedding dimensionality yields substantial performance gains. The GTE-large model (1024-dimensional) outperforms its 768-dimensional counterpart, but optimal results are achieved with Qwen3 models. Specifically, Qwen3-Embedding-8B (4096 dimensions) achieves Top-3 accuracy of 0.571, approximately 16 percentage points higher than GTE-large (1024 dimensions).

This improvement suggests that higher-dimensional embeddings capture richer semantic nuances and contextual information. However, this comes at the cost of increased computational requirements for both embedding generation and vector search operations. Storage requirements scale linearly with dimensionality, and search latency increases due to higher-dimensional distance calculations.

The trade-off between accuracy gains and computational overhead must be carefully considered in production deployments. For latency-sensitive applications, GTE-large (1024-d) combined with reranking may provide an optimal balance, achieving 0.506 Top-3 accuracy with lower computational cost than Qwen3-8B.

### 6.2 Neural Reranking Benefits

Incorporating neural re-rankers provides substantial accuracy improvements. For example, adding the BGE cross-encoder to the GTE-large pipeline increases Acc@3

from 0.412 to 0.506—a gain of approximately 9.4 percentage points. The MiniLM re-ranker similarly enhances top-K accuracy across all K values.

This finding indicates that optimal pipelines should employ semantic search for high-recall initial retrieval, followed by learned re-rankers for precision refinement. The two-stage approach leverages the complementary strengths of bi-encoders (efficiency) and cross-encoders (accuracy).

Interestingly, reranking bridges approximately half the performance gap between 2000-character and 512-character chunking strategies (comparing 0.412→0.506 versus 0.460), suggesting that re-rankers can partially compensate for suboptimal chunking configurations. This finding has practical implications: systems with existing coarse-grained chunking can achieve significant improvements simply by adding a reranking stage.

### 6.3 Chunking Granularity Effects

Finer-grained chunking consistently improves performance across all evaluated models. Reducing chunk size from 2000 to 512 characters raises GTE-large accuracy from 0.412 to 0.460. This improvement likely occurs because smaller chunks:

1. Reduce semantic noise by focusing on specific topics
2. Improve query-passage alignment by creating more targeted retrieval units
3. Enable more precise relevance judgments by reducing the semantic span of each chunk

The service-based semantic chunking (variable sizes) performs comparably to the 512-character recursive split (Acc@3≈0.456), indicating that semantically coherent boundaries are beneficial regardless of the specific segmentation algorithm employed. This suggests that both approaches—fixed fine-grained chunking and semantic boundary detection—achieve similar improvements by reducing noise and maintaining topical focus.

Service-based semantic chunking offers the additional advantage of preserving natural discourse structure, which may be particularly valuable for documents with clear semantic boundaries such as meeting transcripts [1, 10]. For documents with well-defined topical shifts (e.g., different agenda items in meetings), semantic chunking can create more coherent retrieval units without arbitrary mid-sentence splits.

### 6.4 Index Selection

The choice of Milvus index (HNSW versus IVF-Flat) had minor but measurable effects on retrieval quality. HNSW indexing achieved slightly better accuracy in our experiments (Acc@3: 0.460 vs 0.427), likely due to its superior approximation quality for nearest-neighbor search [13].

However, the difference was relatively small compared to the impact of embedding models, chunking strategies, and reranking. This suggests that index optimization should be considered secondary to model selection and pipeline design. For production systems, the choice between HNSW and IVF-Flat should primarily be based on latency requirements, memory constraints, and scalability needs rather than retrieval quality alone.

## 6.5 Pipeline Automation and Reproducibility

Automating the data pipeline through CI/CD integration ensures reproducibility and enables continuous evaluation. By systematically cleaning data, synthesizing QA pairs, and validating results, we obtain a robust test set suitable for longitudinal studies and model comparisons.

The automated pipeline provides several advantages:

- **Consistency**: All models are evaluated on identical test sets
- **Scalability**: New models can be rapidly integrated and tested
- **Version Control**: Results are tracked across model iterations
- **Reproducibility**: Experiments can be reliably replicated

Future work could extend this pipeline with:

- Reference-free evaluation metrics (e.g., LLM-based assessments) [26, 27]
- Integration of user feedback to refine datasets iteratively
- Multi-domain evaluation across diverse document types
- Active learning strategies for selective manual validation

The reference-free approach would enable evaluation without ground-truth labels, potentially expanding evaluation coverage while reducing manual annotation costs.

## 6.6 Limitations and Future Directions

Our study has several limitations:

**Domain Specificity**: Evaluation on US City Council transcripts may not generalize to other domains. Municipal meeting transcripts have specific characteristics (formal language, structured agendas) that may not reflect other document types such as technical manuals, scientific papers, or conversational text.

**Synthetic Ground Truth**: LLM-generated questions may not perfectly represent real user queries. While prior work has validated this approach [25, 26], synthetic questions may exhibit distributional differences from authentic information needs.

**Static Evaluation**: We do not evaluate dynamic aspects such as query reformulation, relevance feedback, or multi-turn interactions. Real search systems often involve iterative refinement that our single-shot evaluation does not capture.

**Computational Cost**: High-dimensional embeddings and neural re-rankers increase latency and resource requirements. Our evaluation focuses on accuracy without considering inference time or throughput constraints relevant to production deployment.

Future research directions include:

**Multi-Domain Evaluation**: Testing on diverse corpora (legal documents, scientific papers, technical manuals, customer support tickets) to assess generalization. Cross-domain evaluation would reveal whether optimal configurations remain consistent across document types.

**Hybrid Metrics**: Combining reference-based and reference-free evaluation approaches to capture different aspects of retrieval quality [26, 27]. Reference-free metrics could assess answer quality, factual accuracy, and coherence.

**Efficiency Optimization**: Investigating quantization, distillation, and approximation techniques to reduce computational overhead [15]. Vector quantization methods could compress embeddings while maintaining retrieval quality.

**User Studies**: Conducting human evaluations to validate synthetic benchmark findings. Direct user feedback would reveal whether metric improvements translate to better user experience.

**Adaptive Chunking**: Developing learned chunking strategies that optimize for retrieval quality. Neural segmentation models could be trained to identify optimal chunk boundaries based on downstream retrieval performance.

**Enhanced Semantic Chunking**: Exploring advanced semantic segmentation techniques including discourse parsing, topic modeling, and rhetorical structure theory. These methods could produce more semantically coherent chunks than simple fixed-size or service-based approaches.

## 7. CONCLUSION

We have presented a comprehensive evaluation of embedding models and pipeline strategies for AI-driven search systems, focusing on municipal meeting transcript retrieval. Our key findings include:

1. **High-dimensional embeddings** (e.g., Qwen3-Embedding-8B with 4096 dimensions) dramatically outperform smaller models, achieving Top-3 accuracy of 0.571 versus 0.412 for GTE-large (1024 dimensions).
2. **Neural reranking** provides substantial gains, with two-stage pipelines (semantic search + cross-encoder) achieving Top-3 accuracy exceeding 0.53.
3. **Fine-grained chunking** (512 characters) and **service-based semantic chunking** consistently improve relevance compared to larger chunks (2000 characters), yielding approximately 5 percentage point accuracy gains. Service-based semantic chunking demonstrates that respecting natural discourse boundaries achieves comparable performance to fixed fine-grained approaches.
4. **Automated pipeline infrastructure** enables reproducible evaluation and continuous integration of model improvements.

Our robust data pipeline—spanning data collection, synthetic QA generation, validation, and CI/CD testing—ensures reliable evaluation and provides a foundation for iterative system improvement. The combination of high-dimensional embeddings, optimized chunking (both fixed fine-grained and semantic), and neural reranking represents the current state-of-the-art for retrieval quality in our benchmark.

The practical implications of our findings are clear: production search systems should prioritize high-dimensional embeddings when computational resources permit, implement two-stage retrieval with neural reranking, and adopt fine-grained chunking strategies. The choice between fixed and semantic chunking depends on document characteristics and available infrastructure.

Future work will expand evaluation to additional domains, explore reference-free evaluation metrics, investigate efficiency optimizations to make high-quality search systems practical for production deployment at scale, and develop more sophisticated semantic chunking approaches that leverage discourse structure and topic modeling.

## ACKNOWLEDGMENTS

The author thanks the Webex Suite AI team at Cisco Systems for their support in developing the evaluation infrastructure and providing computational resources for this research.